# Migration of Celestial Bodies in the Solar System and in Some Exoplanetary Systems

S. I. Ipatov*

*Vernadsky Institute of Geochemistry and Analytical Chemistry, Russian Academy of Sciences, Moscow, Russia*
*\*e-mail: siipatov@hotmail.com*

**Abstract**—A review of the results on the migration of celestial bodies in the Solar System and in some exoplanetary systems is presented. Some problems of planet accumulation and migration of planetesimals, small bodies and dust in the forming and present Solar System are considered. It has been noted that the outer layers of the Earth and Venus could have accumulated similar planetesimals from different areas of the feeding zone of the terrestrial planets. In addition to the theory of coaccretion and the mega-impact and multi-impact models, the formation of the embryos of the Earth and the Moon from a common rarefied condensation with subsequent growth of the main mass of the embryo of the Moon near the Earth is also discussed. Along with the Nice model and the "grand tack" model, a model is considered in which the embryos of Uranus and Neptune increased the semimajor axes of their orbits from values of no more than 10 AU to present values only due to gravitational interactions with planetesimals (without the motions of Jupiter and Saturn entering into resonance). The influence of changes in the semimajor axis of Jupiter's orbit on the formation of the asteroid belt is discussed, as well as the influence of planetesimals from the feeding zone of the giant planets on the formation of bodies beyond the orbit of Neptune. The migration of bodies to the terrestrial planets from different distances from the Sun is considered. It is noted that bodies from the feeding zone of the giant planets and from the outer asteroid belt could deliver to the Earth a quantity of water comparable to the mass of water in the Earth's oceans. The migration of bodies ejected from the Earth is considered. It is noted that about 20% of the ejected bodies that left the Earth's sphere of influence eventually fell back to the Earth. The probabilities of collisions of dust particles with the Earth are usually an order of magnitude greater than the probabilities of collisions of their parent bodies with the Earth. The migration of planetesimals is considered in exoplanetary systems Proxima Centauri and TRAPPIST-1. The amount of water delivered to the inner planet Proxima Centauri b, may have been more than the amount delivered to the Earth. The outer layers of neighboring planets in the TRAPPIST-1 system may contain similar material if there were many planetesimals near their orbits during the late stages of planetary accumulation.

**DOI:** 10.1134/S0038094623600105

## INTRODUCTION

Many authors believe that the protoplanetary disk broke up into rarefied condensations that could collide with each other and eventually turned into solid bodies — planetesimals (e.g., Carerra et al., 2015; Chambers, 2010; Chiang and Youdin, 2010; Cuzzi et al., 2008; 2010; Johansen et al., 2007, 2011, 2012, 2015; Kretke and Levison, 2015; Lorek and Johansen, 2024; Lyra et al., 2008, 2009; Makalkin and Ziglina, 2004; Morbidelli et al., 2009; Rein et al., 2010; Safronov, 1972; Safronov and Vityazev, 1983; Youdin, 2011; Wahlberg and Johansen, 2014). When gas still remained in the protoplanetary disk, the growth of planetary embryos also occurred through the accretion of pebble-type objects (Gurrutxaga et al., 2024; Lambrechts and Johansen, 2012, 2014; Lau et al., 2024; Morbidelli, 2020). This accretion ended when the planetary embryo reached a certain mass (pebble isolation mass). Pebble accretion probably played a greater role in the feeding zone of the giant planets than in the feeding zone of the terrestrial planets. The embryos of the giant planets accreted gas (Safronov, 1972).

This paper briefly discusses some issues of migration of planetesimals and small bodies at the gasless stage in the forming and present Solar System, as well as in the exoplanetary systems Proxima Centauri and TRAPPIST-1. Considerable attention is paid to the results we have obtained. More detailed reviews of the issues under consideration can be found, for example, in the monograph (Ipatov, 2000) and the article (Marov and Ipatov, 2023).

## ACCUMULATION OF TERRESTRIAL PLANETS

In studying the accumulation of terrestrial planets, along with analytical estimates (for example, Lissauer, 1993; Safronov, 1972; Safronov and Vityazev, 1983; Wetherill, 1980), calculations were made of the evolution of the orbits of gravitating bodies that merge during collisions (Carter and Stewart, 2022; Chambers, 2001, 2013; Chambers and Wetherill, 1998; Clement et al., 2018, 2019, 2021; Ipatov, 1987, 1993a, 2000; Izidoro et al., 2014; Joiret et al., 2023; Lykawka and Ito, 2017; Marov and Ipatov, 2023; Morbidelli et al., 2012; Morishima et al., 2010; Nesvorný et al., 2021; O'Brien et al., 2014; Raymond et al., 2004, 2006, 2009; Woo et al., 2022). After the mid-1990s, such calculations took into account the gravitational influence of large bodies by integrating the equations of motion of the bodies. In earlier works, the method of spheres was used to take into account the gravitational influence of bodies. In this method, it was assumed that the bodies moved in unperturbed heliocentric orbits outside the spheres under consideration, and inside the sphere the motion of the bodies was considered within the framework of the two-body problem. Ipatov (1987, 1993a, 2000) considered spheres of action. When selecting pairs of bodies approaching the sphere of action, a "probabilistic" algorithm was first used, in which pairs of bodies approaching to a distance equal to the radius of the sphere of action were selected in proportion to the probability of their approach. Later, an efficient "deterministic" algorithm was developed, in which for a selected pair $i$ and $j$ moment of approaching bodies $t_{ij}$ the closest isolated approach to the radius of the sphere of action is minimal (Ipatov, 1993b, 2000). If the number of bodies in the disk is not small, then when using a deterministic algorithm, the time required for the growth of planetary embryos in the calculations was approximately ten times less than for a "probabilistic" algorithm, in which pairs of approaching bodies were selected in proportion to the probability of approach. Later, numerical integration of the equations of motion yielded planetary growth times close to the results of calculations using the deterministic method. That is, the probabilistic approach is less consistent with real evolution.

Estimates of the time of formation of the bulk of the Earth, about 100 million years, have remained virtually unchanged since the publication of Safronov's monograph (1972). Based on the results of modeling the evolution of disks of initially identical bodies that merge during collisions corresponding to the feeding zone of the terrestrial planets (Ipatov 1987, 1993a, 2000), as well as numerical calculations of the migration of bodies under the influence of planets or their embryos (Ipatov, 2019), the following conclusions were drawn about the accumulation of planets. The inner layers of each terrestrial planet accumulated mainly planetesimals from the vicinity of that planet's orbit. The outer layers of the Earth and Venus could have accumulated similar planetesimals from different regions of the feeding zone of the terrestrial planets. The Earth and Venus could have acquired more than half of their mass in five million years. The formation of a Mars embryo with a mass several times smaller than the mass of Mars as a result of compression of a rarefied condensation can explain the relatively rapid growth of the main mass of Mars. A similar conclusion can be drawn for Mercury. The proportion of planetesimals ejected from the feeding zone of the terrestrial planets into hyperbolic orbits did not exceed 10%. The accumulation of the terrestrial planets was also influenced by bodies migrating from the feeding zone of the giant planets and from the asteroid belt (Clement et al., 2018, 2019, 2021; Morbidelli et al., 2012; Walsh et al., 2011).

## FORMATION OF THE MOON

Several models of the formation of the Moon have been proposed. The theory of coaccretion (Afanasyev and Pechernikova, 2022; Ruskol, 1960, 1963, 1971, 1975) considered the formation of the Moon from a near-Earth swarm of small bodies. The source of the swarm was considered to be the capture of particles from the protoplanetary disk during their collisions ("free-free" and "free-bound"). By "free-free" collisions were meant paired collisions of particles from the Earth's feeding zone, and by "free-bound" collisions were meant collisions of particles with bodies of the proto-lunar swarm around the Earth. A large number of works are devoted to mega-impact models (Barr, 2016; Benz et al., 1986; Canup, 2004, 2012; Canup and Asphaug, 2001; Canup et al., 2013, 2021; Cuk and Stewart, 2012; Cuk et al., 2016; Hartmann and Davis, 1975). In this model, a Mars-sized body collided with the Earth and ejected material from the Earth's mantle formed the Moon. Criticism of the mega-impact model is given in articles (Galimov, 1995, 2011; Galimov et al., 2005) and in a monograph (Galimov and Krivtsov, 2012). In the multi-impact model of the Moon's formation (e.g., Citron et al., 2014; Gorkavyi, 2007, 2023; Ringwood, 1989; Rufu and Aharonson, 2015; Rufu et al., 2017; Svetsov et al., 2012), smaller masses of a larger number of impactors were considered than in the mega-impact model. Salmon and Canup (2012) studied the formation of the Moon from a disk formed by a mega-impact. Galimov and his coauthors (Vasiliev et al., 2011; Galimov, 1995, 2011; Galimov et al., 2005; Galimov and Krivtsov, 2012) considered the model the formation of the embryos of the Earth and the Moon from a single initial gas–dust rarefied condensation, with the subsequent formation and compression of two fragments. In this model the formation of the cores of the Earth and the Moon could not have begun earlier than 50 million years from the origin of the Solar System, dated according to the CAI. In all known works on the for-mation of condensations (see Introduction), the time of formation and compression of condensations was several orders of magnitude less than 50 million years. Ipatov (2018) noted that it is not clear how, in Galimov's model, the evolution of the inner part of the condensation proceeded for a long time, during which iron evaporated, and particles with zero velocities from the outer part of the condensation began to fall onto the nuclei only much later. Due to the zero velocities, these particles should have fallen into the center of the condensation almost instantly (compared to the time it takes for iron to evaporate from the particles in the inner part of the condensation).

Ipatov (2018) showed that the angular momentum of the parent condensation, necessary for the formation of the Earth and Moon embryos, could have been mainly acquired in the collision of two rarefied condensations, in which the parent condensation was formed. Taking into account the subsequent growth of the masses of the Earth and Moon embryos to achieve the current angular momentum of the Earth–Moon system, the total mass of the embryos formed during the compression of the parent condensation could be even less than 0.01 of the Earth's mass. To explain the present proportion of iron in the Moon, the proportion of matter ejected from the Earth's embryo and deposited on the Moon's embryo must have been an order of magnitude greater than the sum of the total mass of planetesimals deposited directly on the Moon's embryo and the initial mass of the Moon's embryo formed from the parent condensation, if the initial embryo contained the same proportion of iron as the planetesimals. Most of the matter that went into the Moon's embryo could have been ejected from the Earth during numerous collisions of planetesimals (and smaller bodies) with the Earth. Unlike the multiimpact model, in the Ipatov (2018) model, when the Moon embryo grew due to bodies ejected from the Earth, the Moon embryo already existed before the start of this growth. The presence of a Moon embryo, formed together with the Earth embryo from a common condensation, can explain why other terrestrial planets do not have a satellite like the Moon. Large bodies also fell on these planets, but they do not have large satellites. When studying the motion of bodies ejected from the Earth, Ipatov (2024) noted that for the embryo of the Moon to grow to its current mass, it had to be located close to the Earth, and not in the current orbit of the Moon.

## ACCUMULATION OF THE GIANT PLANETS

In Safronov's (1972) model, the embryos of the giant planets formed by the accumulation of planetesimals, and then they accreted gas. The total mass of planetesimals exceeded $100 m_E$, where $m_E$ is the mass of the Earth. In articles (Barricelli and Aashamar, 1980; Fernandez and IP, 1981, 1984, 1996; IP, 1989; Ipatov, 1987; Ipatov, 1991, 1993a, 2000) for the feeding zone of giant planets, the results of calculations of the evolution of disks of hundreds of gravitating solid bodies that merge during collisions are presented. Some initial disks included planetesimals and giant planet embryos whose initial orbits were close to those of present planets. When modeling the evolution of disks without planetary embryos (Ipatov, 1987) and with planetary embryos (Ipatov, 1991, 1993a), it was found that the total mass of planetesimals ejected into hyperbolic orbits was an order of magnitude greater than the total mass of planetesimals that became planets. In calculations (Fernandez, Ip, 1984, 1996), bodies were united at approaches of up to 4–8 radii of such bodies, and when taking into account the gravitational influence of planets, spheres smaller than their spheres of action were used, and the mutual gravitational influence of planetesimals was not taken into account. Therefore, in these works the ejection of bodies into hyperbolic orbits turned out to be significantly less than in the article (Ipatov, 1987). In calculations of the evolution of disks of bodies with planets (Ipatov, 1987, 1991, 1993a, 2000), the semimajor axis of Jupiter's orbit decreased over time, while the semimajor axes of the orbits of other giant planets generally increased. During the accumulation of the giant planets, more ice and rocky matter could have entered Jupiter's core and shell than any other planet.

Zharkov and Kozenko (1990) were the first to suggest that the massive embryos of Uranus and Neptune formed near the orbit of Saturn, where they acquired gaseous envelopes. Calculations by Ipatov (1991, 1993a) showed that these embryos could have acquired their present orbits through gravitational interactions with planetesimals. The total mass of planetesimals in the calculations was taken to be equal to $135-150 m_E$. In the considered calculation variants, the initial values of the semimajor axes of the orbits of the embryos of Uranus and Neptune did not exceed 10 AU. During the course of evolution, the semimajor axis of the orbit of Jupiter decreased by $0.005 M_g/m_E$ (Ipatov, 1993a). In my opinion, such a model of the evolution of the orbits of the giant planets can explain the main features of the formation of the Solar System. Later, other models of the evolution of the orbits of giant planets appeared. In the Nice model (Clement et al., 2018, 2019; Gomes et al., 2005; Morbidelli et al., 2005, 2007, 2010; Tsiganis et al., 2005) Jupiter and Saturn entered into resonance, which caused an abrupt change in the semimajor axes of the orbits of Uranus and Neptune. In (Gomes et al., 2005) this abrupt change occurred after 880 million years, when Jupiter and Saturn entered a 1 : 2 resonance. Note that in the calculations of Ipatov (1993a, 2000), due to the gravitational influence of planetesimals, the main changes in the semimajor axes of the orbits of Uranus and Neptune from less than 10 AU to present values occurred over a period of about 10 million years. Therefore, these planets could have acquired their present orbits long before Jupiter and Saturn entered into the 1 : 2 resonance. In (Morbidelli et al., 2007), the jump in the semimajor axes of the orbits of Uranus and Neptune occurred after 2 million years due to Jupiter and Saturn entering the 3 : 5 resonance.

In the "grand tack" model (Jacobson and Morbidelli, 2014; O'Brien et al., 2014; Rubie et al., 2015; Walsh et al., 2011), Jupiter was initially driven by gas and moved toward the Sun up to 1.5 AU. After the formation of massive Saturn and the dissipation of gas, Jupiter began to move together with Saturn back from the Sun, being in the 2 : 3 resonance with Saturn. As a result of this migration, Jupiter "cleaned" the asteroid belt, reduced the amount of material in the feeding zone of Mars, and contributed to the delivery of water to the forming terrestrial planets.

## FORMATION OF THE ASTEROID AND TRANS-NEPTUNE BELTS

The decrease in the semimajor axis of Jupiter's orbit, discussed above, led to a change in the positions of resonances in the asteroid belt and contributed to the cleaning of the asteroid belt. (Safronov and Ziglina, 1991; Torbett and Smoluchowski, 1980). For the 5 : 2 resonance with Jupiter, an example of clearing the resonance zone is given in (Ipatov, 1992). It was shown that the boundaries of the 5 : 2 Kirkwood gap coincide with the boundaries of the region of initial values of the semimajor axes and eccentricities of the orbits of asteroids, at which, during the course of evolution, the orbit of Mars is achieved and the bodies can leave the gap.

During the accumulation of the giant planets, the total mass of planetesimals, which for some time moved beyond the orbit of Neptune, was equal to tens of Earth masses. Five years before the discovery of the first (after Pluto) trans-Neptunian object, based on the results of calculations of the formation of the giant planets, Ipatov (1987) suggested that in addition to trans-Neptunian objects formed further than 30 AU from the Sun and moving in weakly eccentric orbits, objects formed in the zone of giant planets move in this zone in strongly eccentric orbits. These objects are now called "scattered disc objects."

For solid-body accumulation of trans-Neptunian objects from small planetesimals, it is necessary that the accumulation occur at small (~0.001) eccentricities and a massive belt (tens of Earth masses). According to calculations (Ipatov, 1999, 2001), due to the gravitational influence of forming giant planets, trans-Neptunian objects (TNOs) and migrating planetesimals, such small eccentricities could not exist for the time required for the solid accumulation of trans-Neptunian objects with a diameter $d > 100$ km. Therefore, trans-Neptunian objects with a diameter $d \geq 100$ km, moving in not very eccentric orbits, could have formed by the compression of large, rarefied dust condensations (with $a > 30$ AU), and not by accretion of smaller planetesimals. Probably some planetesimals with a diameter $d \sim 100-1000$ km in the feeding zone of the giant planets and even some large main belt asteroids could also have formed directly due to compression of rarefied dust concentrations (Morbidelli et al., 2009). Some smaller objects (TNOs, planetesimals, asteroids) may have been fragments of larger objects, while other such objects may have formed directly through the compression of condensations.

The fraction of small bodies with satellites is 0.1 among main-belt asteroids, 0.15 for near-Earth asteroids, about 0.3 for classical trans-Neptunian objects, and 0.1 for other trans-Neptunian objects. The article (Ipatov, 2017a) shows that the angular velocities of condensations used in the calculations (Nesvorný et al., 2010) as initial data in modeling the compression of rarefied condensations (consisting of dust and/or objects with a diameter less than 1 m), leading to the formation of trans-Neptunian satellite systems, could have been obtained in collisions of condensations whose radii are comparable to their Hill radii. The heliocentric orbits of the colliding condensations could have been close to circular. The model of the formation of the satellite system, considering the collision of two condensations, is consistent with observations that about 40% of binary objects discovered in the trans-Neptunian belt have negative angular momentum relative to their centers of mass (Ipatov, 2017b). The ratio of the tangential component of the collision velocity of condensations to the parabolic velocity on the surface of the condensation is inversely proportional to the distance from the Sun and inversely proportional to the cube root of the mass of the condensation. Therefore, the coagulation of condensations-preplanetesimals is more likely for more massive condensations and for condensations that were located at a greater distance from the Sun (in particular, more likely for condensations in the trans-Neptunian belt than in the asteroid belt).

## MIGRATION OF SMALL BODIES FROM VARIOUS REGIONS OF THE SOLAR SYSTEM TO THE TERRESTRIAL PLANETS

Water and volatiles are important for the origin and evolution of life in the Solar System and in extrasolar systems (Marov, 2023). Earth's ocean water and its deuterium to hydrogen D/H ratio may result from mixing of water from multiple exogenous and endogenous sources with high and low D/H ratios. The problem of migration of celestial bodies in the solar system is also important for understanding the formation and evolution of the solar system. Endogenous sources of water could include direct adsorption of hydrogen from nebular gas into magma melts, followed by reaction of hydrogen $H_2$ with iron oxide FeO and the accumulation of water by particles of the protoplanetary disk before the beginning of gas dissipation in the inner part of the early Solar System, which could have increased the D/H ratio of deuterium to hydrogen in the Earth's oceans (Genda and Icoma, 2008). They also include the accumulation of water by particles in the protoplanetary disk before the beginning of gas dissipation in the inner part of the early Solar System (Drake and Campins, 2006; Muralidharan et al., 2008). Exogenous sources included migration of bodies from the outer part of the main asteroid belt (Ipatov, 2021; Lunine et al., 2003; Morbidelli et al., 2000, 2012; O'Brien et al., 2014; Petit et al., 2001; Raymond et al., 2004) and migration of planetesimals from beyond Jupiter's orbit (Ipatov, 2010, 2020; Ipatov and Mather, 2003, 2004a, 2004b, 2006; Levison et al., 2001; Marov and Ipatov, 2018, 2023; Morbidelli et al., 2000). A number of studies have considered the outer asteroid belt as the main source of water on Earth. Below, we discuss exogenous water sources. Many bodies collided with the terrestrial planets and the Moon during the Late Heavy Bombardment (LHB). Various estimates of this period lie within the range from 4.5 to 3.5 billion years ago. The sources of the LHB were discussed, in particular, in the article (Bottke and Norman, 2017). The article (Ipatov, 1999) estimated that up to 20% of near-Earth objects with a diameter $d > 1$ km could come from the trans-Neptunian belt.

The papers (Ipatov and Mather, 2003, 2004a, 2004b) investigated the evolution of the orbits of ~30000 Jupiter-crossing objects (JCOs) with an initial period of $P < 20$ years. The gravitational influence of seven planets (Venus–Neptune) was taken into account. Resonant asteroids and trans-Neptunian objects were also considered. The symplectic algorithm of the SWIFT package was mainly used to integrate the equations of motion (Levison and Duncan, 1994). For the problems considered, the results of calculations by this method were close to the results of calculations by the method (Bulirsh and Stoer, 1996). The average dynamic lifetime of a JCO is about 100000 years, and the average time for an object initially crossing Jupiter's orbit to move in an Earth-crossing orbit was about 30000 years. However, among the 30000 JCOs considered, there were several objects that moved into orbits lying inside the orbit of Jupiter and moved along them for tens of millions of years. The probability of such an object colliding with a terrestrial planet may be greater than that of 10000 other objects that have crossed Jupiter's orbit. Although only a small fraction of the migrated objects had semimajor axes of their orbits $a < 2$ AU, the average time that objects initially crossing Jupiter's orbit spend in orbits with $a < 2$ AU was comparable to the time at $a = 3$ AU. The probability of a collision of an object crossing Jupiter's orbit with the Earth was about $4 \times 10^{-6}$–$4 \times 10^{-5}$ (interval for different groups of objects).

In the article (Marov and Ipatov, 2018) in the series of "JS" calculations, the present orbits and masses of the terrestrial planets, Jupiter and Saturn were considered. In the $JS_{01}$ series the masses of the terrestrial planets were 10 times less than their present values (in a number of cosmogonic models it is believed that Jupiter and Saturn were almost formed when the masses of the terrestrial planets were still far from their present values). In the JN and $JN_{01}$ series additionally, Uranus and Neptune in their current orbits were considered. In four series of calculations of JS, $JS_{01}$, JN, and $JN_{01}$ the semimajor axes $a$ of the initial planetesimal orbits varied from $a_{\min} = 4.5$ to $a_{\max} = 12$ AU, and the number of planetesimals with a semimajor axis of the orbit close to $a$, was proportional $a^{1/2}$. The eccentricities of the initial orbits of the planetesimals were 0.3 (such were the average eccentricities of the orbits of the planetesimals at the final stages of accumulation of the terrestrial planets), and the inclinations of their orbits were 0.15 rad. The collision probabilities of the planetesimals considered (from the zones of Jupiter and Saturn) with the Earth were of the order of $2 \times 10^{-6}$ (Marov and Ipatov, 2018). About 30% of the water delivered from the feeding zone of Jupiter and Saturn could have been delivered during the growth of the Earth embryo to $0.5 m_E$.

In a number of calculations (Ipatov, 2020, 2021, 2024c; Marov and Ipatov, 2023), the initial values $a_o$ of the semimajor axes of the orbits of planetesimals (bodies) varied from $a_{\min}$ to $a_{\min} + d_a$, their initial eccentricities were equal to $e_o$, and the initial inclinations were equal to $e_o/2$ rad. In some calculations the values $a_{\min}$ varied in 2.5-AU increments from 5 to 40 AU and $d_a = 2.5$ AU, and $e_o = 0.05$ or $e_o = 0.3$. In other calculations the values of $a_{\min}$ varied in 0.1-AU increments from 3 to 4.9 AU and $d_a = 0.1$ AU, and $e_o = 0.02$ or $e_o = 0.15$. The gravitational influence of the planets (from Venus to Neptune) and the Sun was taken into account. The SWIFT integration package was used to integrate the equations of motion. The probabilities of collisions of bodies with planets and the Moon were calculated on the basis of arrays of orbital elements of migrated bodies, since these probabilities are small.

Values $p_E$ of the probability of a collision of a body with the Earth were, on average, smaller for larger initial values of the orbital semimajor axes $a_o$ at $5 \le a_o \le 40$ AU (Ipatov, 2020). Probability values $p_E$ of collisions of planetesimals with the Earth are about $10^{-6}$ at a distance of about 15–40 AU and ~$10^{-5}$ at a distance of about 4–10 AU. For some values of $a_{\min}$ and $e_o$, the values of $p_E$ calculated for 250 planetesimals, can differ by hundreds of times for calculations with almost identical initial orbits. This difference is due to the fact that one in thousands of planetesimals could have been in an Earth-crossing orbit for millions of years.

Bodies that were initially at different distances from the Sun reached the Earth at different times $t$. At $3 \le a_{\min} \le 3.5$ AU and $e_o \le 0.15$ individual bodies could fall to the Earth and the Moon after several hundred mil-lion years. For example, for $a_{min} = 3.3$ AU and $e_o = 0.02$, the value of $p_E = 4 \times 10^{-5}$ at $0.5 \leq t \leq 0.8$ million years (time of the Late Heavy Bombardment) and $p_E = 6 \times 10^{-6}$ at $2 \leq t \leq 2.5$ million years. For $a_{min} = 3.2$ AU and $e_o = 0.15$, it was found that $p_E = 0.015$ at $0.5 \leq t \leq 1$ million years, and $p_E = 6 \times 10^{-4}$ at $1 \leq t \leq 2$ million years. The outer asteroid belt zone may have been one of the sources of the "late heavy bombardment." Most impacts of bodies, originally located at 4 to 5 AU from the Sun, onto the Earth occurred during the first 10 million years. Bodies that initially crossed Jupiter's orbit may have come to Earth's orbit mostly within the first million years (after the formation of massive Jupiter). The time of migration of bodies from the feeding zone of Uranus and Neptune depended on when large embryos of these planets appeared in this zone. According to (Ipatov, 1993a), the main changes in the orbital elements of the embryos of the giant planets occurred in no more than 10 million years. Most bodies with $5 < a_{min} < 30$ AU fell on the Earth over 20 million years. At $a_{min} > 20$ AU the values of $p_E$ could have increased slightly even after hundreds of millions of years, and individual bodies could remain in elliptical orbits and after a time equal to the age of the Solar System.

The total mass of the Earth's oceans is about $2 \times 10^{-4} m_E$. With $p_E = 4 \times 10^{-6}$ and with a total mass of planetesimals $m_\Sigma$ at an initial distance of 5 to 10 AU from the Sun equal to $100 m_E$, the total mass $m_{\Sigma E}$ of planetesimals that fell to the Earth amounted to $4 \times 10^{-4} m_E$. For the zone 10–40 AU with $p_E = 1.5 \times 10^{-6}$ and $m_\Sigma = 100 m_E$, we have $m_{\Sigma E} = 1.5 \times 10^{-4} m_E$. For the zone 3–4 AU with $p_E = 10^{-3}$ and $m_\Sigma = 10 m_E$ we have $m_{\Sigma E} = 0.01 m_E$. The estimated amount of ice in comets does not exceed 33%. However, some authors believe that the primordial planetesimals may have contained more ice (~50%) than present comets. The above estimates indicate that the total mass of water ice delivered to the Earth from beyond Jupiter's orbit could have been comparable to the mass of Earth's oceans. Although bodies in the outer asteroid belt had a smaller total mass and probably contained less ice (~10%) than bodies beyond the orbit of Jupiter, due to the much higher probability of collisions with the Earth, bodies coming from the outer asteroid belt zone could bring to the Earth no less water than bodies from the zone of giant planets. Due to the decrease in the semimajor axis of Jupiter's orbit, which ejected planetesimals into hyperbolic orbits, the zone in the asteroid belt with a higher probability of delivering planetesimals to the Earth shifted and over time covered a larger number of planetesimals. Some of the water was lost during collisions of planetesimals with planets, and especially with the Moon. Consequently, the amount of water that entered the terrestrial planets and the Moon could have been less than the amount of water delivered to these celestial objects.

The total mass of water delivered to Venus and Mercury from beyond Jupiter's orbit, calculated per planetary mass, was approximately the same as for the Earth, and a similar mass of water delivered to Mars, calculated per unit of planetary mass, was approximately 2–3 times greater than for the Earth (Marov and Ipatov, 2018). In absolute terms, the mass of water delivered to Mars was 3–5 times less than the mass of water delivered to Earth. The results support the existence of ancient oceans on Mars and Venus, possibly partially preserved at depth (Mars) or lost during evolution (Venus).

The mass of water delivered to the Moon from beyond Jupiter's orbit could have been less than that for the Earth by no more than 20 times. For planetesimals in the feeding zone of the terrestrial planets, the ratio $r_{EM}$ of the number of planetesimals that collided with the Earth compared to the number of collisions with the Moon generally varied from 20 to 40. Initially, planetesimals that were more distant from the Earth's orbit came to it from more eccentric orbits, and for them the ratio $r_{EM}$ was smaller than that of closer orbits with small eccentricities. In 80% of the variants of calculating the migration of bodies at $3 \leq a_o \leq 5$ AU, we obtained $16.4 \leq r_{EM} \leq 17.4$. In other calculation variants, at $a_o \geq 3$ AU the ratio $r_{EM}$ could be in the range from 14.6 to 17.9. The fraction of water evaporated during impacts with the Moon is higher at higher impact velocities and, accordingly, at lower values of $r_{EM}$.

Using the values of $r_{EM}$ and the formulas given in (Marov and Ipatov, 2021), we estimated the characteristic velocities of collisions of bodies with the Earth and the Moon for some cases. For planetesimals near the Earth's orbit (with $0.9 \leq a_o \leq 1.1$ AU) the characteristic velocities of their collisions with the Moon were from 8 to 10 km/s, with the Earth were from 13 to 15 km/s. For planetesimals that came from other parts of the feeding zone of the terrestrial planets (at $20 \leq r_{EM} \leq 40$), the interval of characteristic velocities of collisions of planetesimals with the Earth was mainly in the range from 13 to 19 km/s, and with the Moon from 8 to 16 km/s. For most bodies with $a_o \geq 3$ AU (at $16.4 \leq r_{EM} \leq 17.4$), a similar range of characteristic collision velocities was 23–26 km/s for the Earth and 20–23 km/s for the Moon. However, the range of collision velocities of all bodies from this zone (at $14.6 \leq r_{EM} \leq 17.9$) was wider: from 22 to 39 km/s for the Earth and from 19 to 38 km/s for the Moon. The above velocities are greater than the parabolic velocities on the surfaces of these celestial bodies, which allows some bodies ejected during collisions to get heliocentric orbits.

The characteristic velocities of collisions of planetesimals, initially located relatively close to the orbit of the Earth embryo, with the Earth and Moon embryos, whose masses were ten times smaller than the present masses of these celestial objects, were generally within the range of 7–8 km/s for the Earth embryo and from 5 to 6 km/s for the Moon embryo. For planetesimals that came from more distant (from the Earth's orbit) regions of the feeding zone of the terrestrial planets, the characteristic velocities were from 9 to 11 km/s for collisions with the Earth's embryo and from 7 to 10 km/s for collisions with the Moon's embryo.

The approaches of small bodies to the Earth are associated with problems of the asteroid-comet hazard (Shustov, 2011; Emelyanenko and Shustov, 2013) and crater formation. A comparison was made (Ipatov et al., 2020) of the distribution of lunar crater diameters in the Oceanus Procellarum region with an age of less than 1.1 billion years with estimates of the number of craters made based on the number of near-Earth objects (NEOs) and the characteristic times that passed before their collisions with the Moon. This comparison suggests that the number of NEOs may have increased following the catastrophic destruction (e.g. 160 million years ago (Bottke et al., 2007)) of a large main-belt asteroid. However, taking into account the destruction of some old craters and the changing orbital distribution of NEOs over time may lead to the conclusion that the average number of NEOs over the last billion years may have been close to the current value. The dependences of the ratio of the crater depth to its diameter on the diameter were studied for lunar craters on the seas and on highlands for craters with an age of less than 1.1 billion years (Feoktistova and Ipatov, 2021). In particular, it was noted that with the same diameter, craters on the lunar seas are deeper than on the continents, with a crater diameter less than 40–50 km. For larger diameters, craters on the continents are deeper.

## MIGRATION OF BODIES EJECTED FROM THE EARTH

During the accumulation stage of the Earth and during the late heavy bombardment, large bodies could collide with the planets. These collisions could have resulted in the ejection of matter from the planets. In the articles (Gladman et al., 2005; Reyes-Ruiz et al., 2012), the motion of bodies ejected from the Earth during collisions of impactor bodies with the Earth was studied over a time interval of 30 thousand years. In these calculations, the initial velocities of the ejected bodies considered were perpendicular to the Earth's surface. It is known that the values of the ejection angle $i_{ej}$ are mainly located between 20° and 55°, especially between 40° and 50° (Shuvalov and Trubetskaya, 2011). In the articles (Ipatov, 2024a, 2025) the motion of bodies ejected from the Earth was studied during the dynamic lifetime $T_{end}$ of all bodies, which usually was about 200–400 million years. In each variant of the calculations, the motion of 250 bodies ejected from the Earth was studied at fixed values of the ejection angle, $i_{ej}$ (measured from the plane of the surface), the ejection velocity $v_{ej}$ and the integration step over time $t_s$ and for a fixed emission point. The gravitational influence of the Sun and all eight planets was taken into account. Bodies that collided with planets or the Sun, or reached 2000 AU from the Sun, were excluded from the integration. To integrate the equations of motion, the symplectic algorithm from the SWIFT package (Levison and Duncan, 1994) was used. The time step under consideration $t_s$ equaled 1, 2, 5 or 10 days, and the calculation results were compared for different $t_s$. In most calculation options, the step $t_s$ was equal to 5 days. The probabilities of collisions of bodies with the Moon were calculated based on arrays of orbital elements of migrated bodies (stored in 500-year increments) similarly to (Ipatov and Mather, 2003, 2004a, 2004b, Ipatov, 2019). Six opposite ejection points on the Earth's surface are considered. In most calculations, the bodies started from the surface of the Earth. In different variants, the ejection angle $i_{ej}$ values were 15°, 30°, 45°, 60°, 89° or 90°. The ejection velocity from Earth $v_{ej}$ varied from 11.22 to 20 km/s (mainly equaled to 11.22, 11.5, 12, 16.4 and 20 km/s).

The average value of the probability $p_E$ of the collision of a body with the Earth depends on the distribution of bodies by $v_{ej}$ and $i_{ej}$. At ejection velocities $v_{ej} \leq 11.3$ km/s, i.e., slightly greater than the parabolic velocity, most of the ejected bodies fell to the Earth. At the ejection velocity $v_{ej}$, equal to 11.5 and 12 km/s, values of $p_E$ did not differ much for different starting points on the Earth's surface and were around 0.2–0.3. The total number of bodies delivered to the Earth and Venus probably did not differ much. In general, the ratio $p_V/p_E$ of the probability $p_V$ of a body colliding with Venus to the probability $p_E$ of a collision with the Earth was smaller at lower ejection velocities.

About 2–6% and 1% of bodies ejected from Earth could have hit Mercury and Mars in $T = 10$ million years. At $T = T_{end}$ (for the entire time interval considered) the probabilities of collisions of bodies with Mercury and Mars were in the ranges of 0.02–0.08 and 0–0.025, respectively. In most calculations, the probability of a body colliding with the Sun was about 0.05–0.2 with $T = 10$ million years, and with $T = T_{end}$ could reach 0.5. The value of the probability $p_{ej}$ of the ejection of bodies into hyperbolic orbits when $T = T_{end}$ exceeded the values at $T = 10$ million years, usually no more than 2 times. The values of $p_{ej}$ in various variants of calculations were in the range of 0.016–0.064 with $T = T_{end}$, 30º ≤ $i_{ej}$ ≤ 60º and 11.5 ≤ $v_{esc}$ ≤ 12 km/s. With $v_{ej} = 16.4$ km/s, the value of $p_{ej}$ could reach 0.8, and at $v_{ej} = 20$ km/s and ejected from a point in the direction of the Earth's motion, all bodies could be ejected into hyperbolic orbits.

The values of the probability of collisions of bodies with the Moon were of the order of 0.01–0.02 at $v_{ej} = 11.3$ km/s and 0.005–0.008 at $v_{ej} = 16.4$ km/s. The average collision velocities of ejected bodies with the Earth and the Moon are greater at greater ejection velocities. With ejection velocities of 11.3, 11.5, 12, 14, and 16.4 km/s, the collision velocities were approximately 13, 14–15, 14–16, 14–20, and 14–25 km/s, respectively, for the Earth and were within 7–8, 10–12, 10–16, and 11–20 km/s for the Moon.

## DUST MIGRATION IN THE SOLAR SYSTEM AND THE FORMATION OF THE ZODIAC CLOUD

Particles formed during asteroid collisions and ejected from comets during sublimation of the icy matrix of the nucleus are the main source of interplanetary dust. When the *Deep Impact* spacecraft's impact module collided with comet Tempel 1 (A'Hearn et al., 2005), a cavity containing gas and dust was excavated (Ipatov and A'Hearn, 2011). The amount of material contained in dust particles and small meteoroids and falling daily on the Earth's atmosphere is from 30 to 180 tons. The migration of dust particles has been considered by a number of authors (Gorkavyi et al., 1997; Ipatov 2010; Ipatov and Mather, 2006; Ipatov et al., 2004; Liou et al., 1995, 1996; Marov and Ipatov, 2005; Moro-Martin and Malhotra, 2002; Reach et al., 1997).

The evolution of the orbits of 20 thousand dust particles was studied in (Ipatov et al., 2004; Ipatov and Mather, 2006) until the particles left the Solar System or collided with the Sun. The BULSTO method (Bulirsh and Stoer, 1996) was used for integration. The gravitational influence of all planets, radiation pressure, the Poynting–Robertson effect, and the solar wind were taken into account. A wide range of particle sizes was considered (from 1 μm to several millimeters). The calculations were performed for β values (the ratio of the radiation pressure force to the gravitational force) equal to 0.0001, 0.0002, 0.0004, 0.001, 0.002, 0.004, 0.005, 0.01, 0.05, 0.1, 0.2, 0.25, and 0.4. For silicate particles with a density of 2.5 g/cm$^3$, such values of β correspond to diameters $d$ of particles of 4700, 2400, 1200, 470, 240, 120, 47, 9.4, 4.7, 2.4, 1.9, and 1.2 μm, respectively. For the same β for water ice, the diameter $d$ is 2.5 times more than for silicate particles. For cometary and asteroidal dust particles, the probability $p_E$ of collisions of particles with the Earth had a maximum (~0.001–0.005) with $0.002 \leq \beta \leq 0.01$, i.e., with $d \sim 100$ μm (this value $d$ is in agreement with observational data). These values of the probability $p_E$ for particles, were generally (except for comet 2P) an order of magnitude larger than for the parent comets. Asteroid particles are shown not to dominate at distances from the Sun of $R > 3$ AU, and a significant portion of the particles at a distance of 3–7 AU are of cometary origin. Dust may have been more efficient than bodies in delivering organic matter to planets. This is because the dust particles were not subjected to intense heating as they passed through the atmosphere (since they had a higher surface-to-mass ratio).

Ipatov et al. (2008) compared velocities determined from the magnesium line in the spectrum of zodiacal dust, obtained from WHAM observations, with estimates based on models for migrating dust particles generated by various small bodies. Based on these studies, it was concluded that the fractions of asteroid particles, cometary particles formed within the orbit of Jupiter, and particles formed beyond the orbit of Jupiter may be of the order of 1/3 each, with a possible deviation from 1/3 by 0.1–0.2. The average eccentricity of the orbits of zodiacal particles located at distances of 1–2 AU from the Sun, which better matches the WHAM observations, is between 0.2 and 0.5, with a more likely value of about 0.3.

## MIGRATION OF BODIES FROM THE FEEDING ZONE OF THE PLANET PROXIMA CENTAURI $c$

Currently, more than five thousand exoplanets have already been discovered (Marov and Shevchenko, 2020, 2022). In the exoplanet system Proxima Centauri, the mass of the star is 0.12 solar masses, and the planet $c$ ($a_c = 1.489$ AU, $e_c = 0.04$, $m_c = 7m_E$) is located behind the ice line. Schwarz et al. (2018) studied the motion of exocomets in the Proxima Centauri system under the distribution of inclinations $i$ of initial orbits of exocomets in the form of the Oort cloud with $0° \leq i \leq 180°$.

In the articles (Ipatov, 2023a, 2023b, 2023c) calculations of the migration of planetesimals from the feeding zone of an exoplanet c were carried out, including to the inner exoplanet $b$ ($a_b = 0.04857$ AU, $e_b = 0.11$, $m_b = 1.17m_E$), which may be located in the habitable zone. Calculations were also carried out for a smaller mass of the planet $c$. In each calculation variant, the initial values of the semimajor axes $a_o$ of the orbits of planetesimals were in the range from $a_{min}$ to $a_{max} = a_{min} + 0.1$ AU. Considered values of $a_{min}$ ranged from 0.9 to 2.2 AU. The number of planetesimals with $a_o$ was proportional $a_o^{1/2}$, i.e., the surface density was proportional $a_o^{-1/2}$. The initial eccentricities of the planetesimals' orbits were $e_o = 0.02$ or $e_o = 0.15$. The initial orbital inclinations of the planetesimals were $e_o/2$ rad. The calculations took into account the gravitational influence of the star and exoplanets c and b. The symplectic algorithm from the SWIFT package was used to integrate the equations of motion. In the main series of calculations, bodies that collided with planets or the star, or reached the Hill sphere of the star (1200 AU) were excluded from the integration.

It was shown that the total mass of planetesimals ejected into hyperbolic orbits exceeded the mass of planetesimals that entered into the planets. Although only one in hundreds of planetesimals reached the orbits of the inner planets, the probability of its collision with these planets could be as high as 1. Probabil-ity of a planetesimal colliding with planet *b*, averaged over all planetesimals considered, was $\sim 10^{-4}$–$10^{-3}$ (Ipatov, 2023c). When averaging over all the planetesimals under consideration that migrated from the vicinity orbits of exoplanet Proxima Centauri *c*, the probability of collision of planetesimals with exoplanets b or d there is greater than the probability of a collision with the Earth of a planetesimal that has migrated from the feeding zone of the giant planets of the Solar System. The latter probability per planetesimal is usually less than $10^{-5}$. The total mass of bodies from the feeding zone of planet *c*, delivered to the planet *b*, was estimated to be $0.002 m_E$ and $0.015 m_E$ at $e_o = 0.15$ and $e_o = 0.02$, respectively. The probability of a dust particle with a diameter of ~100 microns falling from the vicinity of the orbit of the planet Proxima Centauri *c* onto the planet *b* was found to be equal to 0.2 (Ipatov, 2023d), which is significantly higher than the similar probability for planetesimals. A lot of icy material and volatiles could have been delivered to exoplanets Proxima Centauri *b* and *d*.

In (Ipatov, 2023a) the sizes of the feeding zone of planet *c* are considered. Hundreds of millions of years later, some planetesimals could still be moving in elliptical orbits within the feeding zone of planet *c*, which was mainly cleared of planetesimals. Often such planetesimals could move in some resonances with the planet, for example, in resonances of 1 : 1 (like Jupiter's Trojans), 5 : 4 and 3 : 4 (Ipatov, 2023a). There were more such remaining planetesimals at low orbital eccentricities. Some planetesimals, moving for a long time (1–2 million years) in chaotic orbits, fell into resonances of 5 : 2 and 3 : 10 with planet *c* and moved in them for at least tens of millions of years. For some (usually resonant) subregions $a_o$, located outside the main feeding zone of planet *c*, planetesimals could be ejected into hyperbolic orbits or could collide with planets.

The inclinations *i* of the orbits of more than 80% of planetesimals that reached 500 or 1200 AU from the star did not exceed 10°. With the current mass of planet *c* the fraction of such planetesimals at $i > 20°$, on average, for all calculation variants, did not exceed 1% (Ipatov, 2023b). Strongly inclined orbits of bodies in the outer part of the Hill sphere of the star Proxima Centauri can occur mainly only for bodies that entered the Hill sphere from the outside. The radius of the Hill sphere of the star Proxima Centauri (1200 AU) is an order of magnitude smaller than the radius of the outer boundary of the Hill cloud in the Solar System and is two orders of magnitude smaller than the radius of the Hill sphere of the Sun. Therefore, it is difficult to expect the existence of such a massive analog of the Oort cloud around this star as around the Sun.

## MOTIONS OF PLANETESIMALS IN THE TRAPPIST-1 EXOPLANETARY SYSTEM

The TRAPPIST-1 exoplanet system consists of a star with a mass equal to 0.0898 solar masses and seven planets (ranging from *b* to *h*) with masses from 0.33 to $1.37 m_E$. The semimajor axes of the planets' orbits range from 0.012 to 0.062 AU. According to (Payne and Kaltenegger, 2024), the planets *e*, *f*, and *g* may have liquid water. In each variant of the calculations (Ipatov, 2024c), a star, seven planets, and planetesimals were considered, with the initial values of the semimajor axes of the orbits of the planetesimals varied around the semimajor axis of the orbit of one of the planets, the initial eccentricities of their orbits were equal to $e_o = 0.02$ or $e_o = 0.15$, and their initial inclinations were equal to $e_o/2$ rad.

Time of evolution of disks around planets *b*, *c*, *d*, *e*, *f*, *g*, and h took values from 11 thousand to 64 million years. More than half of the planetesimals from disks around the orbits of six planets *b*–*g* collided with the planets in less than 1000 years, and for the disks *b*–*d* even after 250 years. No more than 3% of planetesimals were ejected into hyperbolic orbits. For disks originally located near the orbits of planets *b* and *c*, there was no ejection of planetesimals. There were no collisions of planetesimals with the parent star in the considered calculation variants.

The results of calculations showed (Ipatov, 2024c) that, like Earth and Venus, several planets in the TRAPPIST-1 exoplanet system accumulated planetesimals that were initially located at the same distance from the star. The fraction of planetesimals that collided with the host planet (compared to collisions with all planets) generally decreased with the increasing time interval considered. In each calculation variant, there was at least one planet for which the number of planetesimal collisions exceeded 25% of the number of planetesimal collisions with the host planet. The outer layers of neighboring planets in the TRAPPIST-1 system may contain similar material if there were many planetesimals near their orbits during the late stages of planetary accumulation.

## FUNDING

The article was prepared within the framework of the state assignment to the Vernadsky Institute of Geochemistry and Analytical Chemistry of the Russian Academy of Sciences.

## CONFLICT OF INTEREST

The author of this work declares that he has no conflicts of interest.